\documentstyle[mprocl]{article} 
\input{psfig} 
\def\Journal#1#2#3#4{{#1} {\bf #2}, #3 (#4)} 
\def\NPB{{\em Nucl. Phys.} B} 
 
\def\PRL{\em Phys. Rev. Lett.} 
\def\PRD{{\em Phys. Rev.} D}

\begin{document} 

\hbox to\hsize{%
  \hfil\vbox{%
       \hbox{gr-qc/9711024}%
       \hbox{BUTP-97/27}%
       \hbox{November 6, 1997}%
       }}
\vspace{5mm}
 
\title{MASS INFLATION INSIDE NON--ABELIAN BLACK HOLES
\footnote{To appear in the proceedings of the Eighth Marcel
Grossmann Meeting, Jerusalem, June 1997}
} 
 
\author{ P. BREITENLOHNER$~{^\S}$, G. LAVRELASHVILI$~{^\ddagger}$ 
\footnote{On leave of absence from Tbilisi Mathematical Institute,  
GE-380093 Tbilisi, Georgia} 
, and  D. MAISON$~{^\S}$ } 
 
\address{Max-Planck-Institut f\"ur Physik, F\"ohringer Ring 6$~{^\S}$\\  
80805 Munich, Germany \\ 
Institute for Theoretical Physics, University of Bern,  
Sidlerstrasse 5 $~{^\ddagger}$ \\ 
CH-3012 Bern, Switzerland} 
 
\maketitle\abstracts{ 
The interior geometry of static, spherically symmetric black 
holes of the Einstein-Yang-Mills-Higgs theory is analyzed. 
It is found that in contrast to the Abelian case  
generically no inner (Cauchy) horizon is formed inside  
non-Abelian black holes. 
Instead the solutions come close to a Cauchy horizon but then undergo an enormous   
growth of the mass function, a phenomenon which can be termed  
`mass inflation' in analogy to what is observed for perturbations of the
Reissner-Nordstr{\o}m solution.  
A significant difference between the theories with and without  
a Higgs field is observed. Without a Higgs field the YM field 
induces repeated cycles of mass 
inflation -- taking the form of violent `explosions' -- interrupted by 
quiescent periods and subsequent approaches to an almost Cauchy horizon. 
With the Higgs field no such cycles occur.   
Besides the generic solutions there are non-generic families with  
a Schwarzschild, Reissner-Nordstr{\o}m and a  
pseudo Reissner-Nordstr{\o}m type singularity at $r=0$.} 
   
\newpage

The discovery of regular, static, spherically symmetric 
solutions of the Einstein--Yang--Mills (EYM) equations  
by Bartnik and McKinnon~\cite{lav-BK} has lead to many surprises  
(see e.g.~\cite{lav-surpr} and ref. therein). 
Especially important is the discovery of `coloured' black holes~\cite{lav-bholes}  
which among other things serve as counter examples to the `No-Hair' conjecture.

Until recently only the region outside the horizon  
of these black holes was studied, 
but now also their interior structure is under investigation
~\cite{lav-DGZ1}$^{\!,\,}$\cite{lav-BLM} and new surprises were found. 
 
%%% 
\begin{figure}[ht] 
\rule{5cm}{0.2mm}\hfill\rule{5cm}{0.2mm} 
\psfig{figure=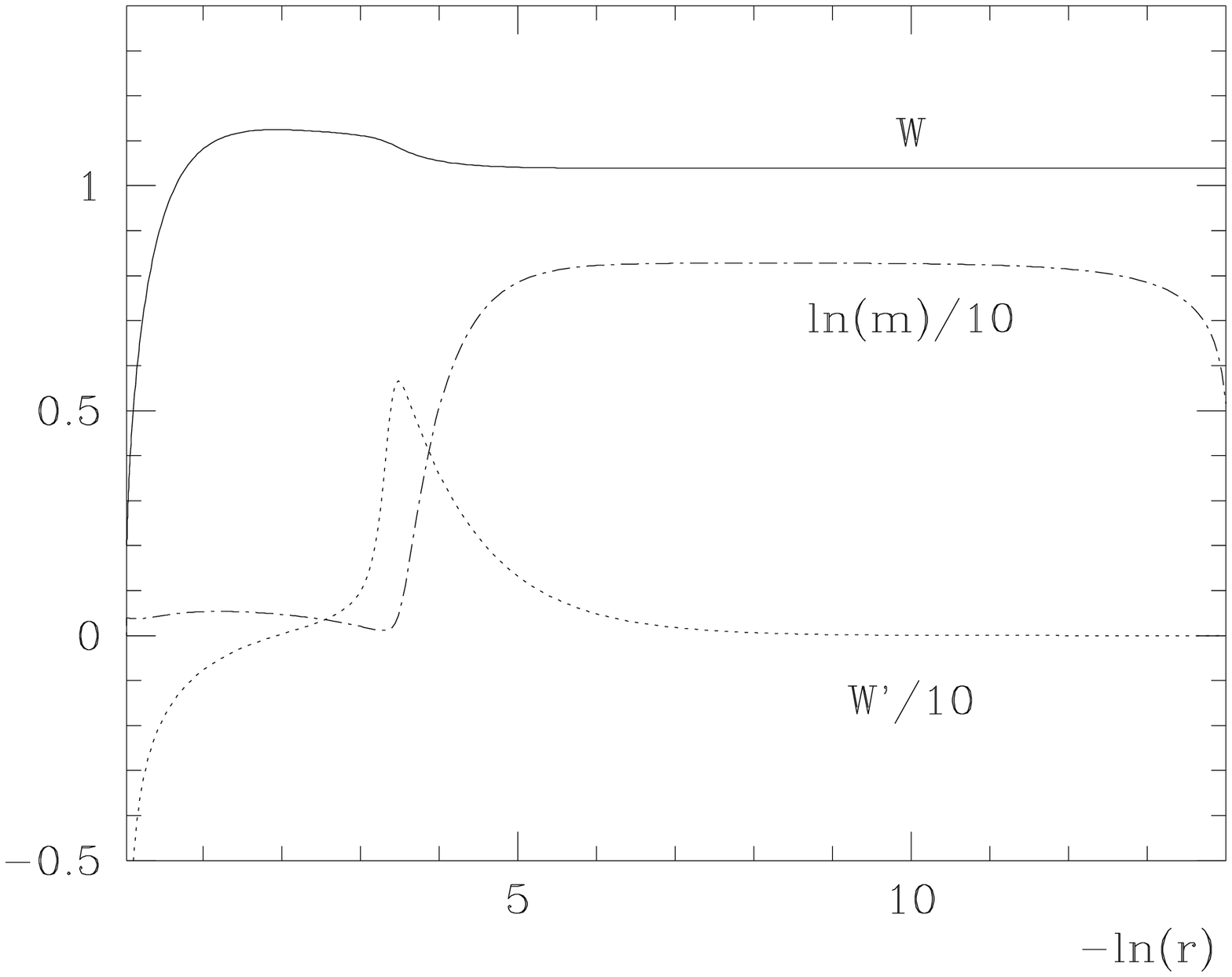,height=1.8in} 
\psfig{figure=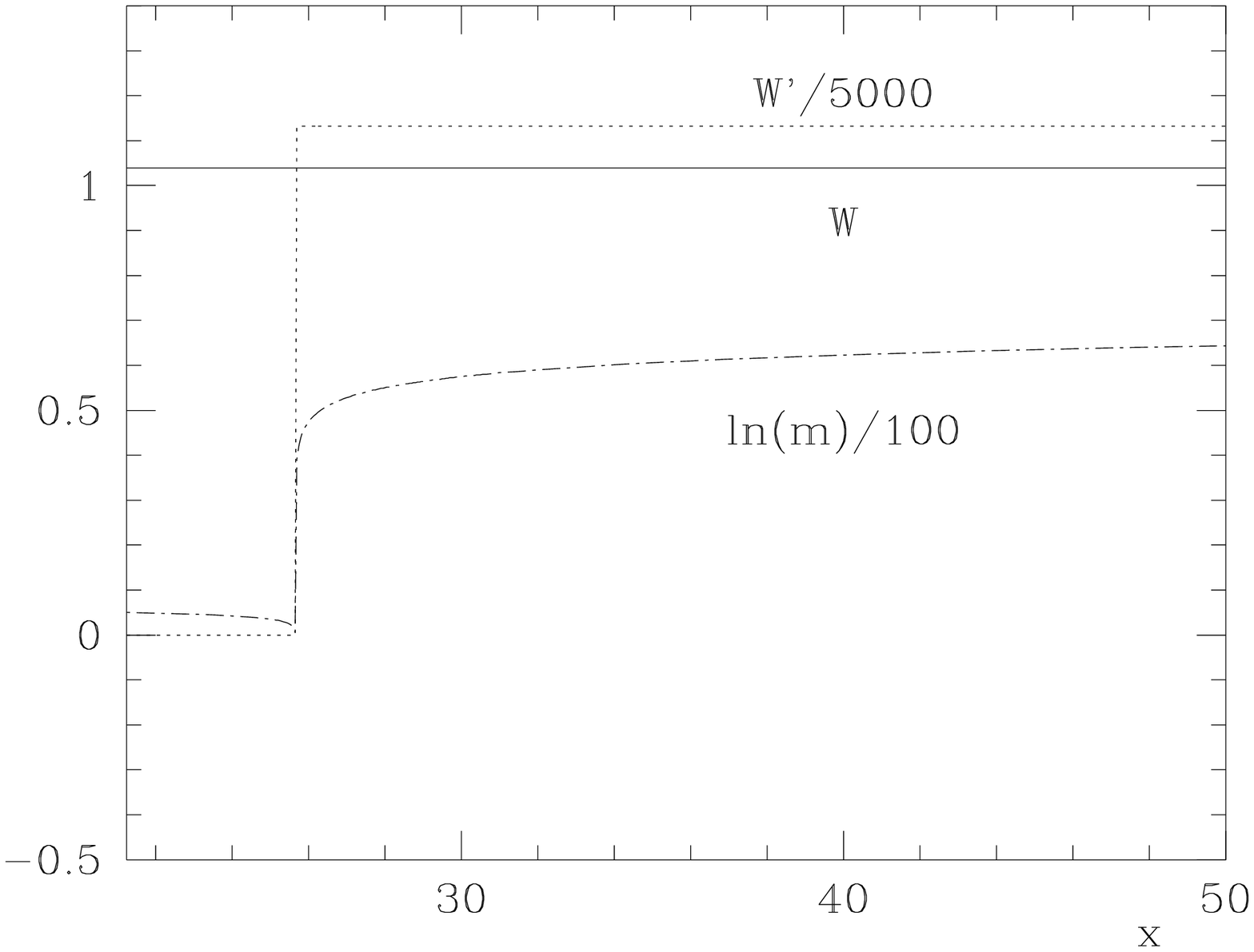,height=1.8in} 
\rule{5cm}{0.2mm}\hfill\rule{5cm}{0.2mm} 
\caption{First two cycles of the generic solution in the 
EYM theory with  parameter values $r_h$=0.97 and $W_h=0.2$.   
For the second cycle a suitably stretched coordinate $x$ is used. 
Different curves represent: the gauge field amplitude $W$,  
its derivative with respect to the Schwarzschild radial coordinate $W'=dW/dr$  
and logarithm of the local mass function $m$. 
\label{fig:eym}} 
\end{figure} 
%%% 
 
The generic black hole solution of the EYM theory was found to be  
oscillatory inside the horizon~\cite{lav-DGZ1}$^{\!,\,}$\cite{lav-BLM}. 
Our numerical results for this case are illustrated in the Fig.~\ref{fig:eym}. 
As one performs numerical integration starting at the horizon  
and integrates towards $r=0$ one observes 
a sudden steep rise of $W'$ and a subsequent exponential growth of the 
mass function $m(r)$ (parametrizing the $g^{rr}$-component of the metric via
$g^{rr}=1-2m(r)/r$). 
This phenomenon can be understood as manifestation of the instability of a
possible 
inner (Cauchy) horizon and is closely related to the `usual'  
mass inflation phenomenon observed for perturbations of the RN and Kerr black 
holes~\cite{lav-PI}$^{\!,\,}$\cite{lav-Ori1}. 
Within a short interval of $r$ the mass function reaches a plateau and stays  
constant for a `while' until it starts to decrease again.  
When the solution comes close to  an inner horizon,  
the same inflationary process repeats itself 
with an even more violent next `explosion'.   
 
By a suitable fine tuning of the initial data at the horizon 
it is possible to obtain different special solutions. 
Black holes with an inner (Cauchy) horizon -- non-Abelian analogues of the   
Reissner-Nordstr{\o}m (RN) solution (NARN) and two types of 
solutions without inner horizon -- generalization of the Schwarzschild  
solution (NAS) and non-Abelian solutions with a pseudo-RN type singularity (NAPRN) 
were found. 
 
A qualitative understanding of the behaviour of the generic 
solutions was gained using a simplified dynamical system  
described
in~\cite{lav-DGZ1}$^{\!,\,}$\cite{lav-BLM}$^{\!,\,}$\cite{lav-BLM2}.  
The generic behaviour is ruled by a fixed point of this system, which is a
repulsive focal point around which the solutions spiral with growing
amplitude.  
It is possible to obtain an approximate solution leading
to a `plateau -- to -- plateau formula'~\cite{lav-BLM}  
relating quantities at one plateau (before the `explosion') to those  
on the next plateau (after the `explosion').  
 
%%% 
\begin{figure}[ht] 
\rule{5cm}{0.2mm}\hfill\rule{5cm}{0.2mm} 
\psfig{figure=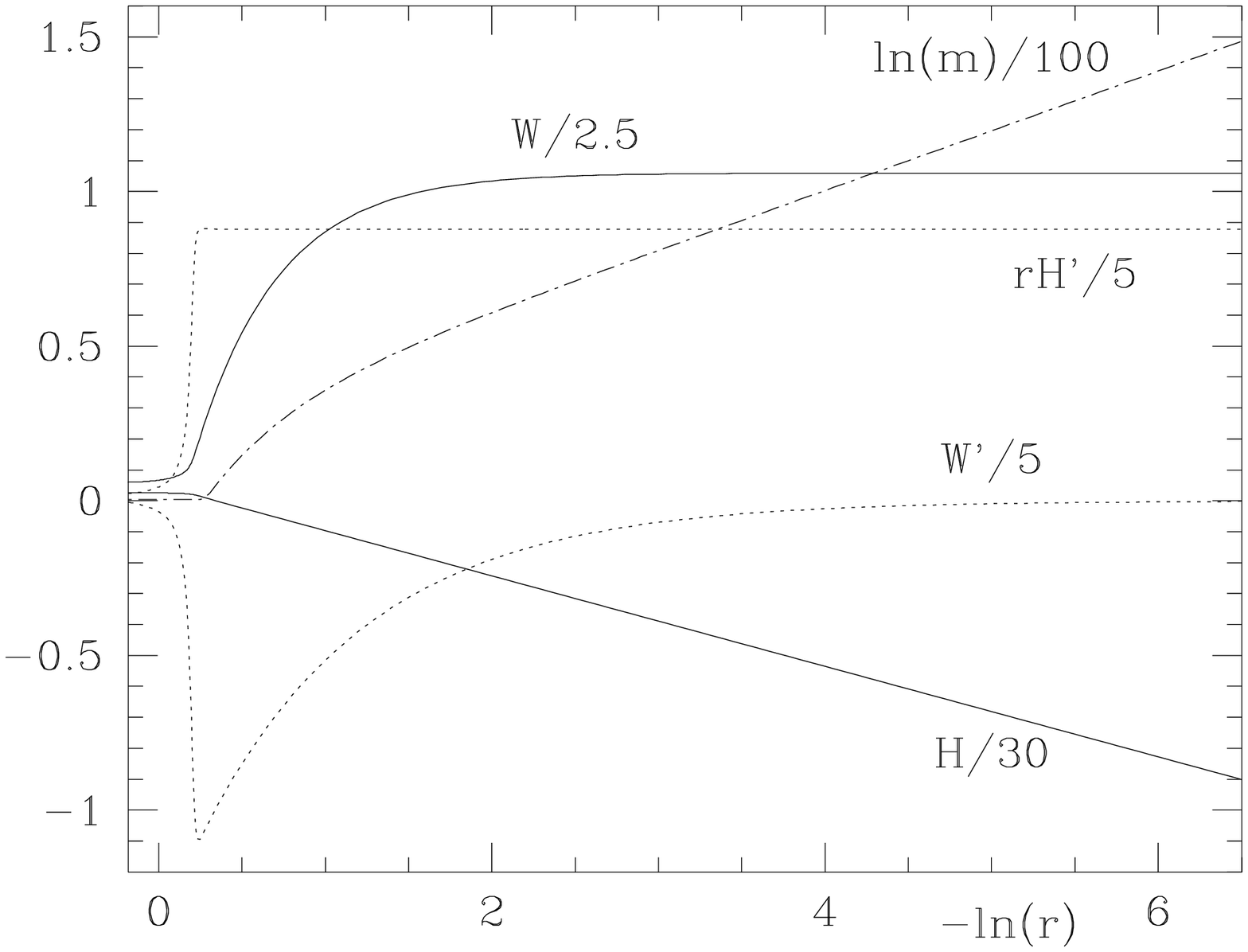,height=1.8in} 
\psfig{figure=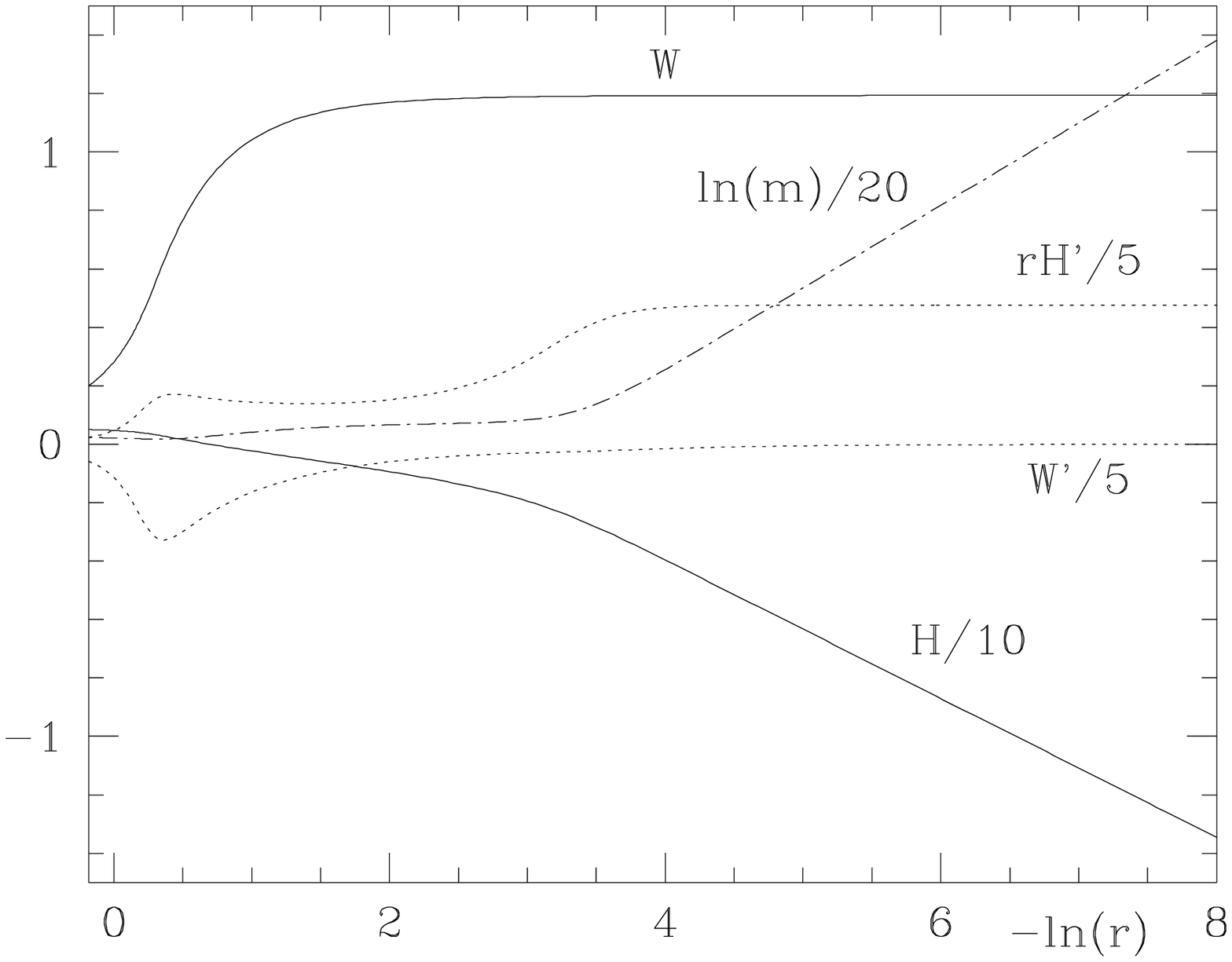,height=1.8in} 
\rule{5cm}{0.2mm}\hfill\rule{5cm}{0.2mm} 
\caption{Two characteristic types of interior solutions with Higgs fields. 
\label{fig:eymh}} 
\end{figure} 
%%% 
 
In order to understand the model dependance of these results 
a theory with an additional Higgs field was 
investigated~\cite{lav-BLM}. 
It was found that after adding the Higgs field no more  
oscillations occur inside the horizon 
\footnote{ 
It is interesting to note that the similar results were obtained  
in the Abelian case in the homogeneous mass inflation model~\cite{lav-Page}.  
}.  
This change in the behaviour of the generic solution is due to the different
fixed points of the corresponding simplified dynamical system.
The focal point disappears and the inflationary behaviour is now ruled by a
a stable attractor 
leading asymptotically to a linear growth of $\ln(m)$ with $\ln(r)$.  

\begin{equation}\label{eq:line} 
\ln(m)=\ln(m_{0}) - z^2_{0}\ln(r),   
\end{equation} 
where $m_{0}$ and $z^2_{0}>1$ are some constants.  
This behaviour is obviously supported by our numerical results
shown in the Fig.~\ref{fig:eymh}. 

The main conclusion is that 
no inner (Cauchy) horizon is formed inside non--Abelian black holes 
in the generic case, instead one obtains a kind of mass inflation. 
Without a Higgs field, i.e.\ for the EYM theory, this mass inflation  
repeats itself in cycles of ever more violent growth. 
Mass inflation is exponential and is assotiated to the instability  
of the inner horizon.  
With the Higgs field there are no such cycles and the mass 
function diverges according to a power law. 
 
\section*{Acknowledgments} 
G.L. is thankful  
to the members of the AEI, Max--Planck--Institut f\"ur Gravitationsphysik  
for the invitation and kind hospitality during his visit to Potsdam,   
when this manuscript was completed. He also thanks the Tomalla Foundation and  
the Swiss National Science Foundation for the financial support.

\end{document}